\documentclass[12pt]{article}
\usepackage{amsmath,amssymb,bbm}
\newcommand{\diag}{\mbox{diag}}
\newcommand{\zz}{\mathbbm{Z}}
\newcommand{\bone}{\mathbbm{1}}
\begin{document}
\title{
\normalsize \hfill UWThPh-2015-25 \\ \hfill IFIC/15-78 \\[5mm]
\Large
Roots of unity and lepton mixing patterns from finite flavour
symmetries 
\thanks{Presented by W.\ Grimus at XXXIX International Conference of
  Theoretical Physics 
  ``Matter to the Deepest,'' Ustro\'n, Poland, September 13--18, 2015}
%
}
\author{R.M. Fonseca$^\mathrm{a}$, W. Grimus$^\mathrm{b}$ \\
\small $^\mathrm{a}$AHEP Group, Instituto de F\'isca Corpuscular,
C.S.I.C./Universitat de Val\`encia \\ 
\small Edificio de Institutos de Paterna,
Apartado 22085, E--46071, Spain 
\\[3mm]
\small $^\mathrm{b}$University of Vienna, Faculty of Physics, Boltzmanngasse 5
\\ 
\small A--1090 Vienna, Austria} 
\date{October 7, 2015}
\maketitle
\begin{abstract}
The classification of lepton mixing matrices from finite residual symmetries
is reviewed, with emphasis on the role of vanishing sums of roots of unity for
the solution of this problem.
\end{abstract}
  
\section{Introduction}
\label{intro}
The $3 \times 3$ mixing matrix or PMNS matrix $U$ in the lepton sector has 
two large and one small mixing angle. It could be that this 
peculiar feature, which is in stark contrast to the CKM matrix in the quark
sector, can be explained through an underlying flavour symmetry.
Many attempts in this direction have been made, but no convincing scenario has
emerged up to now. Some years ago the idea was put forward that the structure
of $U$ is connected with residual symmetries in the charged-lepton and
neutrino mass matrices~\cite{lam}. In this approach, which is completely
independent of any realization of this idea in a model,  
the diagonalization of the mass matrices is effectively replaced by the
diagonalization of the symmetry transformation matrices of the residual
symmetries. Using the notation 
$|U|^2 \equiv \left( \left| U_{ij} \right|^2 \right)$, it turns out that
this approach can either determine $|U|^2$ completely or fix one of its
rows or one of its columns.

In~\cite{fonseca} we have demonstrated that a complete classification of
all possible $|U|^2$, up to independent permutations on $|U|^2$ from the left
and right, can be performed under the following assumptions:
\begin{itemize}
\item
There are three lepton flavours. 
\item
Neutrinos have Majorana nature.
\item
The flavour group $G$ is finite.
\end{itemize}
As a result we have found that there are 
17 sporadic mixing patterns and one infinite series associated with a genuine
three-flavour mixing matrix $U$. All these mixing patterns had been 
found before assuming specific groups~\cite{toorop,hagedorn}. 
Thus our analysis demonstrates   
that there are no possible other mixing pattern, no matter which
finite flavour group $G$ one begins with.

We stress that the 
finiteness of $G$ is an \emph{ad hoc} assumption for the mathematical treatment
of the problem. It is absolutely crucial for the arguments used
in~\cite{fonseca}. 

\section{Residual symmetries}
In order to fix the notation, 
we denote the mass terms of charged leptons and neutrinos by
\begin{equation}
\mathcal{L}_\mathrm{mass} = 
-\bar \ell_L M_\ell \ell_R + 
\frac{1}{2} \nu_L^T C^{-1} M_\nu \nu_L + \mbox{H.c.},
\end{equation}
where the indices $L$ and $R$ indicate the chiralities of the fermion fields
and $C$ is the charge-conjugation matrix. Due to the assumed 
Majorana nature of the neutrinos, we have $M_\nu^T = M_\nu$.
Diagonalization of the mass matrices proceeds via 
\begin{equation}
U_\ell^\dagger M_\ell M_\ell^\dagger U_\ell = 
\diag \left( m_e^2, m_\mu^2, m_\tau^2 \right), \quad
U_\nu^T M_\nu U_\nu = \diag \left( m_1, m_2, m_3 \right),
\end{equation}
leading to the mixing matrix $U = U_\ell^\dagger U_\nu$.

The idea of residual symmetries~\cite{lam} rests on the fact that the 
$\ell_L$ and $\nu_L$ belong to the same gauge doublet, therefore, in a weak
basis they must belong to the same multiplet of the flavour group $G$,
which is broken to the subgroup $G_\ell$ in the charged-lepton sector and
to $G_\nu$ in the neutrino sector. 
Invariance of the mass matrices under the residual groups is formulated as 
\begin{equation}
T \in G_\ell \;\Rightarrow\;
T^\dagger M_\ell M_\ell^\dagger T = M_\ell M_\ell^\dagger,
\quad
S \in G_\nu \;\Rightarrow S^T M_\nu S = M_\nu.
\end{equation}
Since the charged-lepton and neutrino mass spectra are non-degenerate, 
both $G_\ell$ and $G_\nu$ must be Abelian and,
therefore, 
\begin{equation}
G_\ell \subseteq U(1) \times U(1) \times U(1), \quad
G_\nu  \subseteq \zz_2 \times \zz_2 \times \zz_2.
\end{equation}
Consequently, 
all $T \in G_\ell$ together with $M_\ell M_\ell^\dagger$ are simultaneously
diagonalizable, and the same is true for 
all $S \in G_\ell$ and $M_\nu^\dagger M_\nu$.

In essence, the 
diagonalization of $M_\ell M_\ell^\dagger$ is replaced by the diagonalization
of the $T \in G_\ell$ and the 
diagonalization of $M_\nu^\dagger M_\nu$ is replaced by the diagonalization
of the $S \in G_\nu$.

Some remarks are in order.
If a single $T \in G_\ell$ has non-degenerate eigenvalues,
then $U_\ell$ is uniquely determined and $G_\ell \cong \mathbbm{Z}_N$ with a
suitable $N$. In this context it is sufficient that there is one such $T$ in
$G_\ell$. If all $T \in G_\ell$ degenerate, one can show~\cite{fonseca} that
one can confine oneself to two generators $T_1$, $T_2$ of $G_\ell$ and 
$G_\ell \cong K \cong \zz_2 \times \zz_2$ where $K$ is
Klein's four group. With regard to $G_\nu$, one can limit oneself to 
$G_\nu \cong K$ by requiring that all $S \in G_\nu$ have $\det S = 1$.

It is important to realize what the approach 
of residual symmetries achieves and what not. 
Since $U_\ell$ is determined by the diagonalization of the $T$ only up to a
diagonal matrix of phase factors from the right and the analogous statement
holds for $U_\nu$, residual symmetries cannot fix Majorana phases.
Moreover, since we can switch from a representation of $G$ to its complex
conjugate representation, 
the sign of the CKM-type phase in $U$ is not determined either. 
These two statements can be subsumed by saying that with residual symmetries
only 
$|U|^2$ can be determined. Since $|U|^2$ originates in the diagonalization of
representation matrices of $G$, the resulting entries of $|U|^2$ are pure
numbers determined by group theory, independent of the parameters 
of any underlying theory. In addition, this approach does not make any
connection to lepton masses. Therefore, 
$|U|^2$ can only be determined up to independent permutations from
the left and right.

In order to go from groups to mixing matrices $|U|^2$, one first has to
choose a group $G$ which has the subgroup $G_\nu = K$.
Then one has to search for all subgroups $G_\ell$ of $G$ which completely fix
$U_\ell$. Thereafter one has to compute $|U|^2$ for all these subgroups.
Many authors have chosen this approach---see~\cite{fonseca,toorop,hagedorn}
and references therein.

However, a general analysis has to be group-independent. It turns out that the
key to the general analysis is the determination of all possible forms of
$|T| \equiv \left( |T_{ij}| \right)$ with the help of a theorem on vanishing
sums of roots of unity~\cite{conway}, which has already been used in such a
context before in~\cite{grimus}. 

\section{General analysis}
In order to determine the possible forms of $|T|$, 
it is useful to choose a basis where 
$G_\nu = \{ \mathbbm{1}, S_1, S_2, S_3\}$ with 
\begin{equation}
S_1 = \diag\,(1,-1,-1),\; S_2 = \diag\,(-1,1,-1),\; S_3 = S_1 S_2.
\end{equation}
In this basis, $U_\nu = \mathbbm{1}$, $U = U_\ell^\dagger$ and
$UTU^\dagger = \hat T$ is diagonal.

In the following, $P_1$, $P_2$, $P$ are $3 \times 3$ permutation matrices.
There is a series of steps~\cite{fonseca} that leads to possible mixing
patterns: 
\begin{enumerate}
\item
Determination of the five basic forms of $|T|$ up to 
permutations $P_1 |T| P_2$, 
\item
determination of the internal (CKM-type) phase of $T$, 
\item\label{inequ}
finding all inequivalent forms of $|T|$ through $|T| \to |T|P$,
\item
exclusion of two forms of $|T|$ which do not lead to finite groups,
\item
determination of external (Majorana-type) phases of $T$,
\item
computation of possible patterns of $|U|^2$ up to permutations $P_1 |U|^2 P_2$
from the possible matrices $T$.
\end{enumerate}
Here we will only discuss the first step. For the other steps we refer the
reader to~\cite{fonseca}. In relation to step~\ref{inequ} we note that
two matrices $T$, $T'$ are equivalent, \textit{i.e.}\ they lead to the same
$|U|^2$ modulo permutations, if they are related by $T' = V^\dagger T V$ such
that $V$ is a permutation matrix times a diagonal matrix of phase factors. 

In order to perform step~1, we consider the matrices 
$Y^{(ij)} \equiv T^\dagger S_i T S_j$ of $G$. 
With 
\begin{equation}
S_j^{-1} Y^{(ij)} S_j = (Y^{(ij)})^\dagger, \quad \det Y^{(ij)} = 1
\end{equation}
we see that the eigenvalues of $Y^{(ij)}$ must be 
$1,\, \lambda^{(ij)},\, (\lambda^{(ij)})^*$. Due to the finiteness of $G$, 
all $\lambda^{(ij)}$ have to be roots of unity.
Because of 
$\sum\limits_{k=1}^3 S_k = -\bone$, 
\begin{equation}
\sum\limits_{k=1}^3 \mbox{Tr}\,Y^{(kj)} =
\sum\limits_{k=1}^3 \mbox{Tr}\,Y^{(ik)} = 1.
\end{equation}
Written in terms of the eigenvalues, 
these equations give 
\begin{equation}
\sum\limits_{k=1}^3 \left( \lambda^{(kj)} + {\lambda^{(kj)}}^* \right) + 2 = 
\sum\limits_{k=1}^3 \left( \lambda^{(ik)} + {\lambda^{(ik)}}^* \right) + 2 = 0
\;\;\forall\; i,j = 1,2,3.
\end{equation}
It is not difficult to show that there is a relation between $|T_{ij}|$ and the
eigenvalues $\lambda^{(ij)}$:
\begin{equation}\label{abst}
\left| T_{ij} \right|^2 = \frac{1}{2} \left( 1 + \mbox{Re}\,
\lambda^{(ij)} \right).
\end{equation}
Therefore, the generic equation one has to solve is
\begin{equation}\label{generic}
\sum_{k=1}^3 \left( \lambda_k + \lambda_k^* \right) + 2 = 0
\end{equation}
with roots of unity $\lambda_k$. 
Using a theorem of Conway and Jones~\cite{conway}, one can prove that
equation~(\ref{generic}) has, up to reordering and complex conjugation, only
the three solutions 
\begin{equation}
\left( \lambda_1,\, \lambda_2,\, \lambda_3 \right) = 
\left\{
\begin{array}{c}
\left( i,\, \omega,\, \omega \right), \\
\left( \omega,\, \beta,\, \beta^2 \right), \\
\left( -1,\, \lambda,\, -\lambda \right) 
\end{array} \right.
\end{equation}
with
$\omega = e^{2\pi i/3}$, $\beta = e^{2\pi i/5}$ and 
$\lambda = e^{i\vartheta}$ being an arbitrary root of unity.

Any solution $(\lambda_1,\lambda_2,\lambda_3)$ of equation~(\ref{generic}) 
can correspond via equation~(\ref{abst}) to a row or a column of $|T|$. In order
to combine the solutions of equation~(\ref{generic}) to matrices $|T|$, one
must bear in mind that $T$ is unitary, which rules out quite a few
combinations. Up to independent permutations from the left and right, there
are only five forms of $|T|$. Two of these forms do not lead to a finite flavour
group and a third one gives only one sporadic genuine three-flavour mixing
pattern. Thus the two most relevant forms are
\begin{eqnarray}
\left|T\right| &=& \left(\begin{array}{ccc}
0 & \frac{1}{\sqrt{2}} & \frac{1}{\sqrt{2}}\\
\frac{1}{\sqrt{2}} & \frac{1}{2} & \frac{1}{2}\\
\frac{1}{\sqrt{2}} & \frac{1}{2} & \frac{1}{2}
\end{array}\right)
\end{eqnarray}
and 
\begin{eqnarray}
\left|T\right| &=& \left(\begin{array}{ccc}
\frac{1}{2} & \frac{\sqrt{5}-1}{4} & \frac{\sqrt{5}+1}{4}\\
\frac{\sqrt{5}+1}{4} & \frac{1}{2} & \frac{\sqrt{5}-1}{4}\\
\frac{\sqrt{5}-1}{4} & \frac{\sqrt{5}+1}{4} & \frac{1}{2}
\end{array}\right).
\end{eqnarray}
For the further steps in the derivation of the mixing patterns
see~\cite{fonseca}. 

\section{Results}
Confining ourselves to genuine three-flavour mixing patterns, we have found
that, under the assumptions displayed in section~\ref{intro}, 
residual symmetries 
lead to 17 sporadic patterns of $|U|^2$ and one series. Using data on lepton
mixing, it turns out that all sporadic cases are ruled out. 
The mixing pattern of the infinite series is given by 
\begin{equation}\label{series}
\left|U\right|^{2} =\frac{1}{3}\left(\begin{array}{ccc}
1+\textrm{Re}\,\sigma & 1 & 1-\textrm{Re}\,\sigma\\
1+\textrm{Re}\left(\omega\sigma\right) & 1 &
1-\textrm{Re}\left(\omega\sigma\right) \\
1+\textrm{Re}\left(\omega^{2}\sigma\right) & 1 &
1-\textrm{Re}\left(\omega^{2}\sigma\right)
\end{array}\right).
\end{equation}
This $\left|U\right|^{2}$ depends on the parameter 
$\sigma=e^{2i\pi p/n}$, where $p/n$ is a rational number, \emph{i.e.}\ $\sigma$
is a root of unity. 
Clearly, permutation of the rows in equation~(\ref{series})
leads to equivalent mixing patterns due to the freedom in $\sigma$. 
Permutation of the columns, however, leads to three distinct cases in the
usual ordering of charged leptons and neutrino masses, depending in which column
$1/3$ is located. Only the choice displayed in equation~(\ref{series}) for
which 
\begin{equation}\label{12}
\cos^2\theta_{13} \sin^2\theta_{12} = 1/3
\end{equation}
holds is compatible with the data. 
Because of the specific form of $\left|U\right|^{2}$ in
equation~(\ref{series}), the actual parameter which is restricted by the data
on the mixing angles is $\mbox{Re}\,\sigma^6$. 
Using the fit results of~\cite{forero}, 
the 3\,sigma range of $\sin^2\theta_{13}$ translates into 
$-0.69 \lesssim \mbox{Re}\,\sigma^6 \lesssim -0.37$. 
Thus there is indeed a range of $\sigma$ such that the $\left|U\right|^{2}$
of equation~(\ref{series}) is compatible with the data.

In summary, among the  mixing patterns obtained by the approach of
residual symmetries, lepton mixing data single out a unique one-parameter mixing
pattern. The rational number occurring in 
the exponent of the parameter $\sigma$ 
is related to the groups which realize the relevant residual
symmetries. Such groups are subgroups of $SU(3)$ of type~D~\cite{ludl}. 
Apart from the correlations between the mixing angles which can be read off
from the figure in~\cite{fonseca}, a further prediction of
equation~(\ref{series}) is a trivial CKM-type phase.
Therefore, a future measurement of  CP violation in neutrino oscillations
could be a crucial test of this mixing matrix.

\vspace{5mm}

\noindent
\textbf{Acknowledgements:}
The author thanks the Organizers for their hospitality and the nice and
stimulating atmosphere.


\begin{thebibliography}{9}

\bibitem{lam}
C.S.~Lam,
\textit{Phys.\ Rev.\ Lett.} {\bf 101}, 121602 (2008)
\texttt{[arXiv:0804.2622 [hep-ph]]};
\\
C.S.~Lam,
\textit{Phys.\ Rev.} {\bf D78}, 073015 (2008)
\texttt{[arXiv:0809.1185 [hep-ph]]}.

\bibitem{fonseca}
R.M.~Fonseca, W.~Grimus,
\textit{J. High Energy Phys.} {\bf 1409}, 033 (2014) 
\texttt{[arXiv:1405.3678 [hep-ph]]}.

\bibitem{toorop}
R.~de Adelhart Toorop, F.~Feruglio, C.~Hagedorn,
\textit{Nucl.\ Phys.} {\bf B858}, 437 (2012)
\texttt{[arXiv:1112.1340 [hep-ph]]}.

\bibitem{hagedorn}
C.~Hagedorn, A.~Meroni and L.~Vitale,
\textit{J.\ Phys.} {\bf A47}, 055201 (2014)
\texttt{[arXiv:1307.5308 [hep-ph]]}.

\bibitem{conway}
J.~H. Conway and A.~J. Jones,
\textit{Acta Arithmetica} {\bf 30}, 229 (1976).

\bibitem{grimus}
W.~Grimus,
\textit{J.\ Phys.} {\bf G40}, 075008 (2013)
\texttt{[arXiv:1301.0495 [hep-ph]]}.

\bibitem{forero}
D.V.~Forero, M.~Tortola, J.W.F.~Valle,
\textit{Phys.\ Rev.} {\bf D90}, 093006 (2014)
\texttt{[arXiv:1405.7540 [hep-ph]]}.

\bibitem{ludl}
W.~Grimus, P.O.~Ludl,
\textit{J.\ Phys.} {\bf A47}, 075202  (2014)
\texttt{[arXiv:1310.3746 [math-ph]]}.

\end{thebibliography}
\end{document}